# Developing a Dynamic Multi-factor Authentication System for Securing Smart Organizations

Mohsen Saffar
Dept. of Industrial Engineering
K. N. Toosi University of Technology
Tehran, Iran
mohsen.saffar@email.kntu.ac.ir

Hamid Reza Naji*
Dept. of Computer Engineering and Information Technology
Graduate University of Advanced Technology
Kerman, Iran
naji@kgut.ac.ir

***Abstract*—As smart organizations increasingly rely on technologies like AI, big data, and IoT, securing these systems is crucial. This research introduces a dynamic multi-factor authentication (MFA) system within the Zero Trust Architecture (ZTA) framework, which enforces continuous verification. The system randomly selects biometric authentication factors, such as fingerprint, iris, and face recognition using a Markov chain model to ensure adaptability and reduce predictability in security measures.**

**Using CASIA biometric datasets and AES-128 encryption, the system securely manages user data and authentication. Performance analysis shows that the dynamic selection process balances security and efficiency, making it suitable for smart organizations, particularly in e-government contexts. This MFA system strengthens cybersecurity by providing a flexible, real-time defense against evolving threats.**

## I. INTRODUCTION

Electronic government, or e-government, is all about making government services more accessible, efficient, and user-friendly through the use of digital technology, especially the Internet. Instead of waiting in long lines or dealing with piles of paperwork, people can now access services online, from renewing licenses to paying taxes. This shift is designed to simplify processes, reduce red tape, and make government services more available to everyone, no matter where they are or what time it is [1].

Smart organizations play a vital role in making e-government work smoothly. These organizations are the tech-savvy engines behind the scenes, using tools like artificial intelligence, big data, and the Internet of Things (IoT) to help governments operate more efficiently. They enable governments to make better decisions by processing huge amounts of data quickly and securely. They also help streamline everything from managing public resources to handling citizen requests in a faster, smarter way [2].

Security is absolutely essential for smart organizations, especially as they rely more and more on digital tools and data to make things work smoothly. With technologies that becoming central to their operations, these organizations face growing risks from hackers, data leaks, and unauthorized access. Keeping everything secure is not just about protecting sensitive information; it's about building trust with customers, partners, and the public. If security fails, the consequences can be severe, ranging from financial loss to damaged reputations [3, 4].

The Zero Trust security model is crucial for smart organizations, as it ensures that no user, device, or application is trusted by default, requiring continuous verification for access. Unlike traditional security, which assumes internal networks are safe, Zero Trust enforces strict authentication and access control for both internal and external users. It reduces the risk of cyberattacks by limiting access to only what is necessary for each user, enhancing security in cloud environments and remote work setups. By implementing measures like multifactor authentication and real-time monitoring, Zero Trust provides a stronger, more adaptable security framework for smart organizations [5].

In a zero-trust framework, dynamic defense is crucial because it adapts in real-time to changing threats, unlike static defenses that rely on fixed rules. Zero Trust requires continuous verification of users and devices, and dynamic defense supports this by using real-time data and adaptive security measures. This ensures that access controls are always up-to-date and effective against evolving cyber threats, making dynamic defense essential for a robust zero-trust strategy [6].

In this article, we tried to present the dynamic defense mainly under the zero trust structure, which will be a new method for the security of smart organizations in order to achieve the goals of e-government.

## II. REVIEW

**Zero Trust (ZT)** represents a contemporary approach to cybersecurity that shifts the focus from traditional, static network defenses to a model centered on protecting users, assets, and resources regardless of their location or ownership. Rather than assuming trust based on network location, Zero Trust operates on the principle of "never trust, always verify," meaning that every user and device must prove their legitimacy for each access request. This approach is embodied in Zero Trust Architecture (ZTA), which integrates these principles into an organization's cybersecurity framework, encompassing both physical and virtual infrastructure, policies, and access controls [7, 8].

ZTA aims to minimize uncertainties and enforce precise access decisions by requiring continuous authentication and authorization for each transaction or request. Unlike conventional perimeter-based security, which can lead to delays and potential gaps in protection, Zero Trust reduces these issues through rigorous, real-time verification

Published as: M. Saffar and H. R. Naji, "Developing a Dynamic Multi-factor Authentication System for Securing Smart Organizations," 2024 19th Iranian Conference on Intelligent Systems (ICIS), pp. 189-193, 2024. doi: 10.1109/ICIS64839.2024.10887437

processes. It prevents unauthorized internal movements and data breaches by ensuring that every access request is evaluated based on dynamic policies that consider the identity and context of the requester.

Implementing Zero Trust is not about adopting a single technology but rather about following guiding principles that collectively enhance security. Key principles include stringent authentication, granular access controls, and continuous monitoring of all assets to ensure integrity and security. The National Institute of Standards and Technology outlines seven principles for achieving an effective Zero Trust architecture:

1. Define "resources" as any data source or computing service.
2. Ensure all communications are secure, regardless of network location.
3. Restrict access to enterprise resources to individual sessions.
4. Use dynamic policies that factor in the application, client identity, and asset requests to determine access.
5. Continuously monitor and measure the integrity and security of enterprise assets.
6. Enforce strict authentication and authorization before granting resource access.
7. Collect and analyze data about assets, network infrastructure, and communications to improve security.

This approach underscores the importance of a comprehensive and adaptive security strategy in today's smart organizations e- e-government.

In today's digital age, nearly every aspect of modern society has been transformed by technology, making robust authentication processes essential for security. User authentication is a fundamental element of any security system, serving as the primary defense against unauthorized access and impersonation on computing devices. It involves verifying the identity of a user, device, or entity to grant access to resources within a computer system. Authentication methods are typically categorized into three main factors: what the user knows (knowledge factor), what the user possesses (possession factor), and what the user is (inherent factor) [9, 10].

The evolution of authentication systems began with single-factor authentication, progressed to two-factor authentication, and has now advanced to multi-factor authentication (MFA). MFA, which combines multiple types of credentials, offers a higher level of security by providing continuous protection against unauthorized access to critical services and computing devices. As organizations adopt Zero Trust Architecture (ZTA), which emphasizes stringent verification and minimal trust, MFA is increasingly recognized as a crucial component for achieving these security goals [11, 12].

Commonly used MFA methods include:

1. Password protection
2. Token-based authentication
3. Voice biometrics
4. Facial recognition
5. Iris scanning
6. Vein recognition
7. Fingerprint scanning
8. Thermal image recognition
9. Geolocation
10. Behavioral detection
11. Beamforming techniques
12. Occupant Classification Systems (OCS)
13. Electrocardiogram (ECG) recognition
14. Electroencephalography (EEG) recognition
15. DNA recognition

Given that MFA enhances security by requiring multiple forms of verification, it aligns well with the principles of Zero Trust, which seeks to maximize security while minimizing reliance on trust. Therefore, MFA is an effective method for securing digital environments and supporting the zero-trust model.

Historically, static defense mechanisms have been the primary approach to protecting cyber systems from malicious attacks. However, integrating dynamic defense strategies into the design and operational phases of a cybersecurity system is increasingly recognized as essential. Dynamic defense involves continuously evaluating and adjusting all components of an information system, including participants, communication protocols, and data, throughout the system's lifecycle. This approach aims to address the evolving nature of cyber threats by adapting defenses in real-time, rather than relying on static, fixed measures [13-15].

Research indicates that early detection and response account for approximately 95% of the effectiveness in mitigating a cyberattack, while the actual attack phase represents only 5%. Therefore, implementing dynamic defense systems that can intervene early, potentially misleading attackers with false information, is crucial for effective network protection. Among various dynamic defense techniques, Moving Target Defense (MTD) has emerged as a leading strategy in both academic and practical applications. MTD involves altering system parameters or authentication dynamics to enhance resilience against attacks, particularly against sophisticated, stealthy threats [16].

Dynamic authentication, as part of MTD, plays a key role in defending against various types of cyberattacks by continuously adapting to new threat patterns. This dynamic approach is a cornerstone of the zero-trust security model, which emphasizes ongoing trust evaluations and adaptive authorization processes to maintain robust security.

Randomization plays a crucial role in enhancing security defenses, particularly in dynamic environments where threats are constantly changing. Research highlights the effectiveness of dynamic defense mechanisms that incorporate randomness to counteract various cyber threats. MFA, which requires two or more independent credentials, is one such method that boosts security. For that it can use MTD, regularly changing system parameters or authentication processes to protect against sophisticated and stealthy attacks using randomness systems like Markov chains.

Markov chains have proven valuable in various applications, including security protocols. In the realm of multifactor authentication, Markov chains can improve the security and reliability of the authentication process by modeling complex system behaviors and transitions. Additionally, Markov chains are useful for representing and analyzing system dynamics during impersonation attacks, providing a detailed understanding of potential vulnerabilities and behaviors [17].

We used Markov chains to random allocation of authentications of MF as a dynamic method of Zero Trust security system for smart organization. This modeling technique enables a comprehensive approach to enhancing system security in the face of evolving threats.

## III. Material and Methods

This research examines a multi-factor authentication system that incorporates fingerprint recognition, iris recognition, and face recognition. Each of these technologies will be explained further. However the most important feature of the provided system is that it does not require the user to select the authentication factor. Instead, the system randomly selects this element for the user, as we shall explain in the following.

For iris, fingerprint, and face identification in this work, the reliable and extensive CASIA dataset created by the Chinese Academy of Sciences' Institute of Automation was utilized. The CASIA Iris Image Database 4.0 collection, has 54,601 images from 700 individuals, the CASIA Fingerprint Image Database 5.0 collection, has 20,000 fingerprint images from 500 individuals, and the CASIA Face Image Database 5.0 collection, which has 2,500 face images of 500 different people, have all been used for iris recognition, fingerprint recognition, and face recognition. These high-resolution, variable-condition data are gathered, and access to them is typically restricted to authorized researchers.

An SQL database has been utilized to store usernames and extracted features from user's photos. The database has an explicit definition of four columns: username, fingerprint features, iris features, and face features. Fingerprint, iris, and facial data are encrypted using the AES-128 algorithm. Within the authentication procedure, the user initially inputs his username. Next, the presence of the username in the database is verified by using a SQL Query. If the username is present in the database, the authentication procedure proceeds by randomly choosing the initial authentication factor. If the username is absent or wrong, an error message appears, prompting the user to re-enter the username. To choose the authentication element at random, the dynamic authentication system employs a Markov chain. To find the transition probability between various states, a 4x4 transition matrix is first created. The authentication factor is determined by randomly selecting one of ten numbers that are created in a random sequence. Algorithms that are used are sequence generation for authentication factor selection, transition matrix generation, and moving to the next state. Random values are used to create the transfer matrix, which is then normalized for every row. Ultimately, the authentication factor is chosen and returned as output based on this sequence and beginning state. The steps associated with each authentication factor are reviewed below.

First, features from the fingerprint image will be extracted in order to recognize the fingerprint. To verify or deny the user's identity, the data in the second column is decrypted using information taken from the database that was encrypted using AES-128. It is then compared with features extracted from the fingerprint image.

Prior to detecting the iris and extracting data from it, the Haar Cascade method is utilized in order to identify the eye. It is necessary to have the file haarcascade_eye.xml in order to use this approach. This file is utilized for the purpose of identifying eyes in photographs. This file provides a description of the eye features that are utilized in the process of eye recognition. After the eyes of the picture have been identified, the image is then transformed to black and white if it was previously colored. After that, the Gaussian filter is used to the image in order to filter out the noise. Last but not least, the edge detection approach is utilized in order to determine the boundaries of the image. Following the identification of the iris, the iris that has been recognized is next utilized to extract features, such as significant areas. The data in the third column of the database that was encrypted with AES-128 is decrypted and compared with the features that were acquired from the iris image in order to validate or deny the identity of the user.

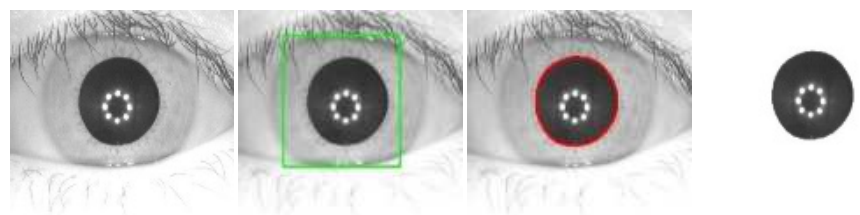

Fig. 1. Stages of eye identification and iris segmentation.

The Haar Cascade method is utilized once more in order to first achieve face recognition, which is a necessary step before actual face recognition can take place. It is necessary to have the file haarcascade_frontalface_default.xml in order to implement this method. This file is utilized for the purpose of recognizing faces in photographs. This file provides a description of the face characteristics that are utilized in the process of training a facial recognition model. Following the completion of facial recognition, the significant characteristics and information of the detected face are extracted. The information contained in the fourth column of the database, which refers to the face and was encrypted using AES-128 encryption, is decoded and compared with the features extracted from the face image in order to either validate or deny the identification of the user.

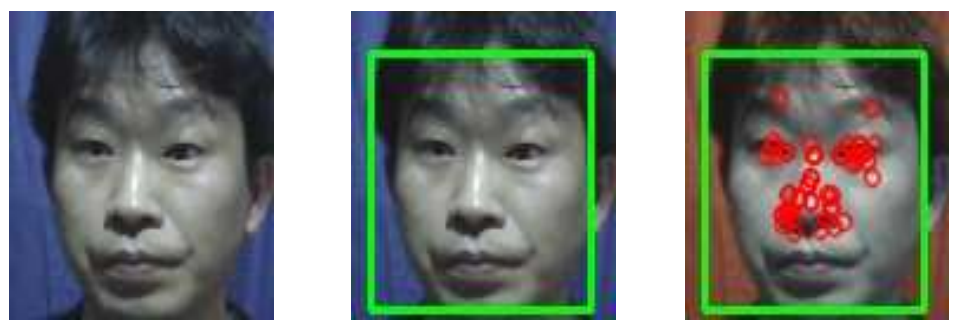

Fig. 2. Stages of face identification.

Following the extraction of distinguishing characteristics from the photographs and the decoding of the information pertaining to the selected agent, the comparison phase is carried out in order to authenticate the user. The Brute Force Matcher (BFMatcher) method is one of the tools that are available in the OpenCV library. It is utilized by this system in order to compare the characteristics of two different images. Using this method, the distance between the features of two photographs is computed, and then the features that are closest to each other are chosen. The Hamming distance is utilized in this investigation for the purpose of comparing each characteristic that was extracted from the photographs. Following successful completion of these steps, the user will be granted access to the system.

Fig. 3 displays the flowchart of the multi-factor authentication system that is implemented.

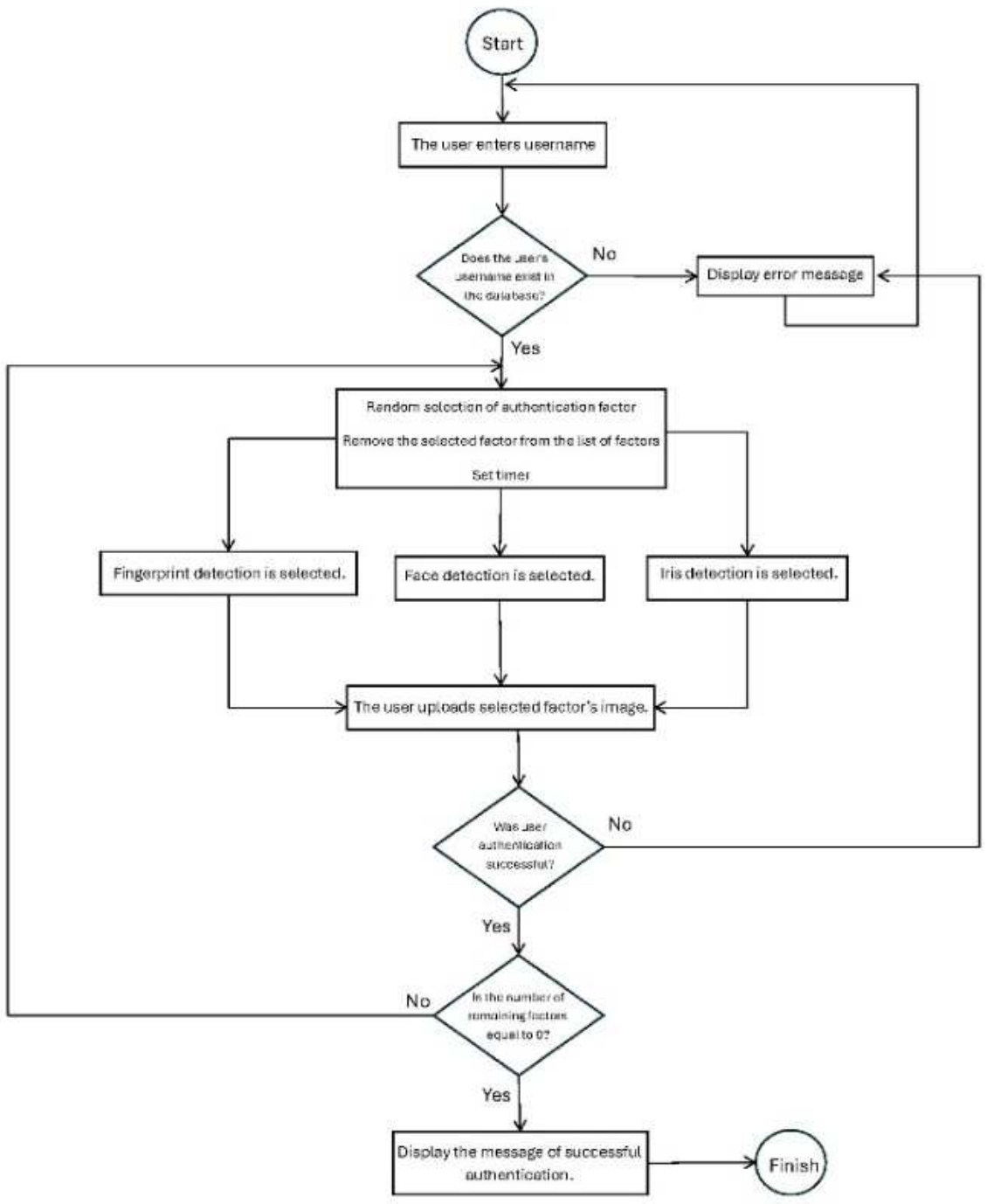

Fig. 3. Flowchart of implemented multi-factor authentication system.

## IV. ANALYSIS OF THE MODEL

The analysis of the outcomes from the multi-factor authentication system that was obtained is provided below.

Multi-factor authentication is a security technique whose implementation might be complicated. It uses multiple techniques to confirm the identity of the user. The performance of this system for evaluation is among its most crucial features. It is not easy to gauge how well user-interactive procedures are performing, though. Thus, performance evaluation in this case is done manually. Table (1-4) lists the average authentication time for each agent individually as well as the average time spent choosing an authentication method. There are 200 samples used in this analysis.
In order to assess the system's overall performance after four tries, Table I. provides the average authentication time.

TABLE I. THE RESULTS OF THE EXECUTION TIME OF THE PROVIDED AUTHENTICATION SYSTEM.

| Authentication Factor | Execution time |
|---|---|
| Fingerprint | 1246 ms |
| Iris | 2867 ms |
| Face | 2943 ms |
| Random Selection | 36 ms |

The analysis of time complexity and performance of the multi-factor authentication system is as follows:

**Iris**: Includes converting the image to grayscale, edge detection, iris segmentation, feature extraction, and descriptor matching. The overall time complexity of this section is $O(k.n)$, where $n$ is the image size and $k$ is the number of keypoints.

**Fingerprint**: Involves feature extraction, data decryption, and matching. The time complexity for this process is $O(k.n)$ where $n$ is the size of the decrypted data and $k$ is the number of keypoints.

**Face**: Similar to fingerprint, it involves image processing and descriptor matching. Its time complexity is also $O(k.n)$.

**Factor Selection Using Markov Model**: Involves generating a transition matrix and moving between states based on probability distributions. The time complexity of this section is $O(n^2 + l.n)$, where $n$ is the number of states and $l$ is the length of the selection sequence.

**Overall System Complexity**: The total time complexity is expressed as $O(n^2 + l.n + T_{Fingerprint} + T_{Face} + T_{Iris})$ where $T$ represents the time complexity of each respective section.

The system's overall complexity depends on the random selection process and the combination of the different factors executed in sequence.

All the items mentioned above are included in TABLE II.

TABLE II. THE TIME COMPLEXITY OF THE PROVIDED AUTHENTICATION SYSTEM.

| Factor | Time complexity |
|---|---|
| Fingerprint | $O(k.n)$ |
| Iris | $O(k.n)$ |
| Face | $O(k.n)$ |
| Random Selection | $O(n^2 + l.n)$ |
| Whole system | $O(n^2 + l.n + T_{Fingerprint} + T_{Face} + T_{Iris})$ |

## V. CONCLUSION

This paper presents a novel dynamic multi-factor authentication (MFA) system included into the Zero Trust Architecture (ZTA) framework, particularly tailored to enhance the security of intelligent organizations. In order to guarantee adaptability and reduce predictability in security measures, the system utilizes a Markov chain model to randomly choose biometric authentication factors, including fingerprint, iris and face recognition. Utilizing the CASIA biometric dataset and AES-128 encryption, the system effectively handles user data and authentication, achieving a harmonious equilibrium between security and efficiency, making it especially well-suited for e-government applications. The results of the performance analysis validate that this method improves cybersecurity by offering versatile and immediate protection against developing threats. This approach is in accordance with the principles of Zero Trust Architecture (ZTA), which prioritize ongoing verification and low reliance, and showcases the efficacy of randomization in reducing complex cyberattacks. In summary, the dynamic MFA system is a strong and reliable solution for intelligent organizations, offering an adaptable and secure approach to authentication in the constantly evolving digital environment.